\documentclass[12pt,letterpaper]{article}

\usepackage{amsmath}
\usepackage{amsfonts}
\usepackage{young}

\usepackage{fancybox}
\usepackage{color}
\usepackage{pstricks}

\usepackage{ifpdf}

\ifpdf
  \usepackage[pdftex]{graphicx}
\else
  \usepackage[dvips]{graphicx}
\fi

\def\be{\begin{equation}}
\def\ee{\end{equation}}
\def\bear{\begin{eqnarray}}
\def\eear{\end{eqnarray}}

\title{Comments on k-Strings at Large N}
\author{Joanna~L.~Karczmarek, Gordon~W.~Semenoff and Shuhang Yang}

\begin{document}

\maketitle

\begin{abstract}
We present a computation of the k-string tension in the large N limit of the two-dimensional lattice
Yang-Mills theory. It is well known that the problems of computing the partition function and the
Wilson loop can be both reduced to a unitary matrix integral which  has a third order phase transition separating
weak and strong coupling. We give an explicit computation of the interaction energy for $k$-strings in the
large $N$ limit when $\tfrac{k}{N}$ is held constant and non-zero.
In this limit, the interaction energy is finite and attractive.
We show that, in the strong coupling phase,
the $k\to N-k$ duality is realized as a first order phase transition. We also show that the lattice $k$-string
tension reduces to the expected Casimir scaling in the continuum limit.

\end{abstract}

K-string tension has been proposed as an interesting probe of the confining phase of Yang-Mills theory
\cite{Strassler:1997ny}\cite{Strassler:1998id}\cite{DelDebbio:2003tk}\cite{DelDebbio:2001sj}\cite{Greensite:2002yn}\cite{Armoni:2003ji}\cite{Armoni:2003nz}\cite{KorthalsAltes:2005ph}\cite{Orland:2006ym}\cite{Bringoltz:2008nd}.  In the dual superconductor picture of confinement,
lines  of electric flux which emanate from a source with color charge are confined to flux tubes,
 the confining strings.  If a colored source has center-charge
 $k$, it is the endpoint of $k$ flux tubes.  There is immediately an interesting question as to whether these flux tubes attract
and perhaps bind together to form a single tube with $k$ units of flux, or whether they repel each other and
 tend to remain as individual vortices, that is whether the dual superconductor is of type I or type II, respectively.

\begin{figure}
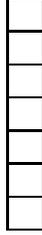

\begin{center}  ${   \begin{Young}
  ~    \cr
     ~       \cr
     ~       \cr
     ~       \cr
     ~       \cr
     ~       \cr
    ~       \cr
    \end{Young}}$
\end{center}
\caption  {The Young tableau corresponding to a completely antisymmetric representation ${\cal A}_k$ of SU(N)
consists of a single column with $k$ boxes.}
\label{f0}
\end{figure}

In Yang-Mills theory, or any gauge theory with only adjoint (or other center-neutral) matter, the
k-string tension $\sigma_k$ is defined as
\begin{equation}
e^{-\sigma_k A} = \frac{1}{{\rm dim} {\cal A}_k}\left<{\rm Tr}_{{\cal A}_k}e^{i\oint_C A_\mu dx^\mu}\right>~,
\label{wilson}\end{equation}
where $C$ is typically a rectangular loop with dimensions $L\times T$, $A=LT$ is the area subtended by that loop and
${\cal A}_k$ is the irreducible representation of the gauge group where the Young tableau consists of a single column of $k$ boxes, depicted in Fig. \ref{f0}.
We are assuming that the gauge group is $SU(N)$.
The physical interpretation of (\ref{wilson}) lies in the idea that, if we introduce a heavy $k$-quark
and a heavy anti-$k$-quark separated a distance $L$ much greater than the confinement scale
into the confining phase of Yang-Mills theory, the energy  attributable to the
gauge theory mediated interaction between them is $E=\sigma_k L$.  The linear growth of the energy with distance
 is a signal of confinement. The coefficient
 $\sigma_k$ is a measure of the strength of the confining interaction.

An intuitive way of thinking about the $k$-string is to imagine that we bring $k$ confining strings close to each other.
If these elementary strings have an attractive interaction,  $\sigma_k < k\sigma_1$,
and the strings can from a bound state which we call the $k$-string.
The $k$-string tension is not easily accessible to perturbation theory and, aside from some supersymmetric models where exact results can be obtained \cite{Douglas:1995nw}, it is most readily studied by numerical simulations of lattice gauge theory.
In this Letter, we will study it in a solvable model, the large $N$ limit of two-dimensional Yang-Mills theory
that was originally formulated and solved by Gross and Witten \cite{Gross:1980he}.  The fundamental and adjoint Wilson loops
in this model have been used as models of deconfinement transitions in finite temperature Yang-Mills theory {Khokhlachev:1979tx}\cite{Dumitru:2003hp}\cite{Pisarski:2005nw}.

If we were to consider an expectation value of a Wilson loop
as in (\ref{wilson}), but taken in some other representation of the
gauge group where the Young tableau also has $k$ boxes,
the expectation value should turn out to be independent
of the representation and equal to that obtained for a k-string
in the completely antisymmetric representation ${\cal A}_k$.
This is because we expect the color charge of the quarks to
 be screened by the gluons which, since they are center charge neutral,
conserve the center charge $k$ but, by combining with the strings of electric flux,
can  alter the precise representation.  This screening is not expected to occur
in two dimensions where there are no dynamical gluons.  It is
also not expected to occur in the large $N$ limit where large $N$
factorization means that mixing of representations is suppressed by
factors of $\tfrac{1}{N}$.
Lack of mixing of representations would lead us to expect that in two
dimensions, or at large $N$, the k-string tension is always given by
the weak coupling result, $\sigma_k = k \sigma_1$.
At a first glance, large $N$ factorization would suggest that the
interaction energy of a $k$-string, $\sigma_k-k\sigma_1$, should be of
order $\tfrac{1}{N^2}$ in the 't Hooft large $N$ expansion, and therefore
vanishingly small in the large $N$ limit.  However, since at infinite $N$
all representations have the same tension, lifting the degeneracy at
finite $N$ can lead to corrections of order
$1/N$, as we will see shortly
\footnote{
In the infinite $N$ 't Hooft limit, and with small values of $k$, all representations with the same
center charge should have the same string tension, $\sigma_k=k\sigma_1$,
that is, they are degenerate.  These representations
are mixed with each other and the degeneracy is resolved by non-planar corrections where the ``interaction''
is of order $\tfrac{1}{N}$. A simple model for this mixing is
the two-state system where the Hamiltonian is
$$
H=\left(\begin{matrix} E_1 & 0 \cr 0 & E_2 \cr \end{matrix}\right) + \left(\begin{matrix} 0 & g/N \cr g/N & 0 \cr \end{matrix}\right)
$$
Corrections to the energy are give
$$
E_1\to E_1 - \frac{g^2}{N^2}\frac{1}{E_2-E_1}+\ldots
~,~E_2\to E_2 + \frac{g^2}{N^2}\frac{1}{E_2-E_1}+\ldots
$$
However, if the leading order is degenerate, $E_1=E_2=E_0$ then
$$
  E_1\to E_0 - \frac{g}{N}+\ldots
~,~E_2 \to E_0 + \frac{g}{N}+\ldots
$$
Generically, corrections are of order $1/N$. (A version of this argument appears in Ref.~\cite{KorthalsAltes:2005ph}.) 
We thank Barak Bringoltz for comments which motivated us to clarify
this argument.}.  There are lattice simulations which suggest that corrections to the $k$-string tension in $3$ dimensions
indeed scale like $\tfrac{1}{N}$, rather than $\tfrac{1}{N^2}$ \cite{Bringoltz:2008nd}.
In addition, the lattice simulations seem to indicate that the splitting of the energies of representations with the
same center charge is identifiable, is also of order $\tfrac{1}{N}$ and is unexpectedly small. 

The $k$-string tension is expected to be symmetric under the replacement of $k$ by $N-k$ so that
\begin{equation}\label{duality}
\sigma_k = \sigma_{N-k}~.
\end{equation}
This is obtained by replacing probe quarks by anti-quarks and vice-versa.
In contrast, weak coupling implies that $\sigma_k \approx k\sigma_1$.
While this fundamental  $k \rightarrow N-k$  symmetry is  compatible with
$k$-strings being weakly coupled
for  $k\ll N$, since $k$ and $N-k$ cannot both be much less than $N$,
in the regime where $N$ is large and $k\sim N$, $k$-strings cannot be weakly interacting.

Continuum Yang-Mills theory in two dimensions is solvable and the $k$-string tension is known exactly,
\begin{equation}\label{continuumformula}
\sigma_k^{D=2}= \lambda C_2({\cal R}_k)~,
\end{equation}
where $\lambda=g^2N$ is the 't Hooft coupling and $g$ is the Yang-Mills theory coupling constant.
$C_2(R)$ is the quadratic Casimir operator of the representation ${\cal R}_k$.
As expected, because there are
no dynamical gluons, the string tension is not independent of representation.
For a completely antisymmetric representation ${\cal A}_k$, it is
\begin{equation}
C_2({\cal A}_k) = \tfrac{k(N-k)}{N}~.
\end{equation}
This expression is symmetric under the replacement
$k\to N-k$.
This $k$-string tension has
$\sigma_k^{D=2}=k\sigma_1^{D=2} - \tfrac{k^2}{N}\sigma_1^{D=2}$, the first term
interpreted as the energy of $k$ non-interacting fundamental strings and  the second
term being the interaction energy.  Thus, the interaction between $k$-strings in
two dimensions seems to be attractive when they are in the antisymmetric representation.
It is also easy to see that for expectation values of a Wilson loop in
 other representations, the interaction energy can be repulsive.

 The {\it Casimir scaling hypothesis} is that the $k$-string tension in higher dimensions remains
 proportional to $C_2({\cal A}_k)$, dependent on the center charge $k$ but, 
 because of screening, otherwise
 independent of the representation of the probe quarks.
This hypothesis is seemingly contradicted by
 the sine-law: an alternative, first principles result found in softly
broken ${\cal N}=2$ supersymmetric Yang-Mills theory \cite{Douglas:1995nw}
 where the $k$-string is a BPS object
 \begin{equation}
 \sigma_k = m\Lambda N\sin\tfrac{\pi k}{N}~.
 \end{equation}
 The sine-law is consistent with results found in MQCD
\cite{Hanany:1997hr} and AdS/CFT \cite{Herzog:2001fq}.
 One interesting difference between the Casimir scaling and
sine-law behaviors is that the corrections to the leading
 order in the large $N$ limit of Casimir scaling are
 of order $1/N$ whereas for the sine-law they are of order $1/N^2$.
In both cases, though, k-strings are strongly interacting when $k \sim N$ as
was discussed above.

In the following, we shall find a third expression for the $k$ string tension
which follows from the large $N$ limit of two dimensional lattice Yang-Mills theory.
Above, we have pointed out two non-generic features of the $k$-string in two
dimensions and at large $N$.  One is the absence
of dynamical gluons in two dimensions which leads to representation dependence,
as is already seen, for example, in the continuum formula (\ref{continuumformula}).
The other is the absence of mixing of representations which happens in the large $N$ limit in
any dimensions, so that
$k$-strings are non-interacting.  In the following, we shall find a way to circumvent
the second shortcoming by studying representations with center charge $k$ where $k\sim N$.
As has been discussed, this is not compatible with weak coupling; mixing of representations,
which would be suppressed by powers of $\tfrac{k}{N}$, is allowed and is of unit magnitude 
when $k\sim N$.
Since the interaction is not suppressed by a power of $\tfrac{1}{N}$,
one question that our computation can answer is whether confining strings continue
to have an attractive interaction so that $k$-strings are stable when the
interaction is strong.  We shall find that it is indeed attractive  and $\sigma_k-k\sigma_1$ is 
negative for all values 
$0<\tfrac{k}{N}<1$.
We shall also confirm that the string tension in the two-dimensional model indeed depends on
the representation and, for the completely symmetric representation with center charge
$k$, the interaction between confining strings is repulsive for all values of $\tfrac{k}{N}>0$.

A Wilson loop in the two dimensional lattice Yang-Mills theory is computed
by the integral
\begin{equation}
W_{\cal R}[C]=
\frac{
\int \prod_\ell[dU_\ell] e^{ \tfrac{N}{\lambda} \sum_P{\rm
Tr}\left(\prod_{\ell\in \delta P} U_\ell + \prod_{\ell\in \delta
P}U^\dagger_\ell\right) }~\frac{1} {{\rm dim}{\cal R}}
{\rm Tr}_{\cal R} \prod_{\ell\in C} U_\ell }{ \int \prod_\ell[dU_\ell] e^{
\tfrac{N}{\lambda} \sum_P{\rm Tr}\left(\prod_{\ell\in \delta P} U_\ell +
\prod_{\ell\in \delta P}U^\dagger_\ell\right) } }
\end{equation}
where $P$ and $\ell$ label plaquettes and links, respectively, of a square
lattice, $U_\ell$ is a unitary matrix residing on the links,
$[dU_\ell]$ is the  invariant Haar integration measure for $SU(N)$, ${\cal
R}$ is an irreducible representation and
${\rm Tr}_{\cal R}$ is the character in that representation. This model
was first solved in the large $N$ limit
by  Gross and Witten \cite{Gross:1980he}.
Using the gauge fixing of Ref.~\cite{Gross:1980he} and the Peter-Weyl theorem $\int[dU]U_{ij}^{\cal R}U_{kl}^{\dagger{\cal R}'}
=\tfrac{\delta_{il}\delta_{jk}\delta_{{\cal R}{\cal R}'}}{{\rm dim}{\cal R}}$, it is possible to reduce this to a one-plaquette unitary
matrix model \cite{Yang}
\begin{align}
W_{\cal R}[C]~&=~\left[w_{\cal R}[C]\right]^{A} ~=~ e^{-\sigma_{\cal R} A}~, \nonumber \\
\sigma_{\cal R}~&=~- \ln\left[~\frac{
\int  [dU ] e^{\tfrac{N}{\lambda} {\rm Tr} \left(U  + U^\dagger \right)}~\frac{1} {{\rm dim}{\cal R}}
\left[{\rm Tr}_{\cal R}  U\right]  }{
\int  [dU ] e^{\tfrac{N}{\lambda} {\rm Tr} \left(U  + U^\dagger \right) }} \right]~,
\label{oneplaquettemodel}\end{align}
where $A$ is the area of the loop -- the number of plaquettes in an area that is bounded by the loop --
and the string tension $\sigma_{{\cal R}}$ is measured in units of inverse lattice spacing squared.
  Note that the Wilson loop in the one plaquette model is normalized by the dimension of
the representation.  This is a result of repeated use of the Peter-Weyl theorem when reducing to the one-plaquette model.  It
has the important consequence that, since in this simple one-matrix model, the dimension of the representation is the upper bound on the expectation value of the character of a matrix in that representation, the string tension is always positive and
the gauge theory interaction is always confining.

The one-plaquette model which computes $\sigma_{\cal R}$ in (\ref{oneplaquettemodel})
is solvable in the large $N$ limit. Gauge invariance
allows one to diagonalize $U={\rm diag}\left(e^{i\phi_1},\ldots,e^{i\phi_N}\right)$ and
the remaining integral over eigenvalues can be done in the saddle-point approximation where the resulting
action is of order $N^2$, so that fluctuations are suppressed by powers of $\tfrac{1}{N^2}$.  The classical configuration of the eigenvalues is determined by minimizing
the action plus a term arising from the integration measure. In all of the cases that we shall consider, if we
seek only the leading order in large $N$, the loop inserted into (\ref{oneplaquettemodel}) can be treated as a probe where
the expectation value is evaluated simply by plugging in the classical distribution of eigenvalues that is determined by the action. This distribution
is characterized by the eigenvalue density, defined as the large $N$ limit of $\rho(\phi)=\tfrac{1}{N}
\sum_i\delta(\phi-\phi_i)$.
The model (\ref{oneplaquettemodel}) has a third order phase transition when $\lambda=2$.  The two phases have
eigenvalue densities \cite{Gross:1980he}
\bear\label{dist1}
\rho(\phi)&=&~ \tfrac{1}{2\pi}\left( 1+\tfrac{2}{\lambda}\cos\phi\right)~,
~\textrm{for}~ \lambda>2~,
\\
\label{dist2}
\rho(\phi)&=&~ \tfrac{2}{\pi\lambda}\cos\tfrac{\phi}{2}\sqrt{\tfrac{\lambda}{2}-\sin^2\tfrac{\phi}{2}}~,
~\textrm{for}~ 0<\lambda<2~,
\eear
 where, in (\ref{dist2}), $|\phi|<2\arcsin\sqrt{\tfrac{2}{\lambda}}$ and on the remainder of the circle $\rho=0$.

 The easiest representations to consider are the totally antisymmetric ${\cal A}_k$ whose Young tableau consists of a single column
 of $k$ boxes and the totally symmetric ${\cal S}_k$ whose Young tableau is a row with $k$ boxes. The analysis of the symmetric representation in a unitary one-matrix model
 was explained in detail in Ref.~\cite{Grignani:2009ua} and we will refer the reader
 there for the  details.  Here, we will concentrate on the antisymmetric representation,
corresponding to a k-string.

 In terms of eigenvalues, the symmetry can be used to order the indices in the trace so that they are non-decreasing
\begin{align}
 {\rm Tr}_{{\cal A}_k}U =\sum_{a_1< \ldots< a_k} e^{i\phi_{a_1}}e^{i\phi_{a_2}}
\ldots e^{i\phi_{a_k}}~.
\end{align}
It is convenient to obtain this expressions from a generating function \cite{Hartnoll:2006is}\cite{Grignani:2009ua}
\begin{align}\label{symmrep}
 {\rm Tr}_{{\cal A}_k}U =
\oint \frac{dt}{2\pi i t^{k+1}}\prod_{a=1}^N\left(1+te^{i\phi_a}\right)~,
\end{align}
where the contour integral projects onto the term in a Taylor
expansion of the integrand which contains $k$ eigenvalues.
The covariant expression is
\begin{align}
\langle {\rm Tr}_{ {\cal A}_k}~U \rangle
~=~ \oint \frac{dt}{2\pi i t^{k+1}}
 \langle \exp\left[{\rm Tr}\ln(1 + tU)\right] \rangle~.
 \label{asym}
\end{align}
 In the large $N$ limit, the traces in the exponents in the above equations
 are replaced by integrals over the eigenvalue densities
(\ref{dist1}) or (\ref{dist2}):\footnote{
The reader might have the concern that the presence of the loop
variable in the path integral, though it does not alter the eigenvalue
distribution to the leading order $N^0$, it will have an effect at
order $1/N$ and a $1/N$ correction in the order $N^2$ part of the
action would contribute a term of order $N$ which competes with the
string tension which we are computing.  To see why this is not a problem,
consider the tension in the large $N$ limit is given by
\begin{equation}\label{565}
\sigma_k =\inf_{(\rho,t)}\left[
N^2S[\rho] + \lambda\int\rho -\lambda-N\int \rho\ln\left(1+te^{i\phi}\right)+k\ln t
\right]
-\inf_{\rho}\left[N^2S[\rho]+\lambda\int \rho -\lambda\right]
\end{equation}
where $S[\rho]$ is the effective action consisting of $S_{\rm eff}$ plus a
contribution from the integration measure.  The saddle-point equations
are
\begin{align}
\frac{\delta S}{\delta\rho} - \frac{1}{N}\ln\left(1+te^{i\phi}\right)+\frac{\lambda}{N^2}=0
~~,~~
\int \rho=1 ~~,~~
\int \rho\frac{te^{i\phi}}{1+te^{i\phi}}=\frac{k}{N}\frac{1}{t}
\label{equationfort}
\end{align}
for the first infimum and
\begin{align}
\frac{\delta S}{\delta\rho} + \frac{\lambda}{N^2}=0
~~,~~
\int \rho=1
\label{equationfort1}
\end{align}
for the second infimum. The eigenvalue density which satisfies
(\ref{equationfort1}) is $\hat\rho_0$. Then the density which satisfies
(\ref{equationfort}) differs from it by a correction of order
$\frac{1}{N}$, $\hat\rho_0+\frac{1}{N}\hat\rho_1$.  However, since
$\hat\rho_0$ satisfies (\ref{equationfort1}), it is easy to see that
if we are interested in $\sigma_k $ only to accuracy of order $N$,
we can simply use $\hat\rho_0$ in the equation which determines
$\hat t$ and, to the same accuracy (where we trust the order $N$ but
not the order $N^0$ contribution), in the expression for
$\sigma_k $ in (\ref{565}). This justifies our use of the ``probe
approximation'' where we use the eigenvalue distribution of the
effective unitary matrix model to compute the generating function.
We note that a similar probe
approximation is made when analyzing the dual objects on the string
theory side of the AdS/CFT correspondence in Ref.\cite{Grignani:2009ua}.}
\begin{align}
\langle {\rm Tr}_{ {\cal A}_k}~U \rangle
~=~
\oint \frac{dt}{2\pi i t^{k+1} }
  \exp\left[N\int d\phi\rho(\phi)\ln(1 + te^{i\phi})-k\ln t\right]~.
 \label{asym1}
\end{align}
When $k$ is large, we can use the saddle-point approximation to evaluate the
integral over $t$ in (\ref{asym}).  Let
$\hat t$ satisfy the saddle-point equation
\be\label{sym4}
R_{ {\cal A}_k}(\hat t) \equiv\int_{-\pi}^\pi d\phi\rho(\phi) \frac{\hat
te^{i\phi}}{1+\hat te^{i\phi}}=\frac{k}{N}~.
\ee
The functions $R_{{\cal A}_k}(t)$ in (\ref{sym4})
are related to the resolvent of the matrix model and are
holomorphic functions of $t$ with cut singularities on the unit circle
determined by the support of $\rho(\phi)$.  With the strong coupling phase (\ref{dist1}), $\lambda>2$,
the cut singularity on the left-hand-side of (\ref{sym4}) occupies the entire unit circle and divides the
plane into two regions, the interior and the exterior of the unit circle. We extend each of these regions to disjoint
Riemann sheets, each of
which covers the entire plane.   We then search for solutions of the saddle-point equations on each of these sheets.
The saddle-point equations for each of the two sheets are
\bear
\tfrac{1}{\lambda}\hat t&=&\tfrac{k}{N}~,  \label{sheet1}\\
1 - \tfrac{1}{\lambda}\frac{1}{\hat t}&=&\tfrac{k}{N}~.  \label{sheet2}
\eear   and the solutions are
\bear\hat t_1&=& \lambda \tfrac{k}{N}\label{sheet11}~,\\
\hat t_2 &=& \tfrac{1}{\lambda}\tfrac{1}{1-k/N}~, \label{sheet22}
\eear
respectively. Using these we find two branches of the string tension, each coming from anti-derivatives by $\hat t$ of
({\ref{sheet1}) and (\ref{sheet2}) respectively.  The possibilities are
\be
\sigma_k  = -k-(N-k)\ln(1-\tfrac{k}{N})+k\ln\lambda    ~~\textrm{and}~~
\sigma_k  =~ -(N-k)+k\ln\tfrac{k}{N}+(N-k)\ln\lambda ~.
\label{sigma-two-poss}
\ee
In principle, to get the correct result, we must choose the one that has the smaller
value of the tension, opening the possibility of a phase transition.  However,
we need to be careful because using a smooth density function $\rho$ in (\ref{asym1})
might lead to the wrong result when $k/N$ is large enough.  A smooth $\rho(\phi)$
is an approximation to $N^{-1}\sum_a \delta(\phi-\phi_a)$.  To accurately
obtain a $k^{\textrm{th}}$ Taylor expansion coefficient of $t$ in the argument of the exponent
in equation (\ref{asym1}), we need to know the $k^{\textrm{th}}$ Fourier coefficient
of $\phi$ in $\rho(\phi)$.  When $k/N$ (which we are asuming to be of order 1 as $N\rightarrow \infty$)
is larger than the inverse of the density of eigenvalues in some region, replacing
the discrete sum over eigenvalues with the continuum density might not lead to the correct answer,
since the wavelength of the fourier mode we are interested in will be shorter than the
inter-eigenvalue spacing.  Equation (\ref{asym1}) thus should not be trusted for $k/N$ large enough.
However, if we analiticaly continue the $t$-integral onto the second sheet, we are effectively
exchanging $k$ and $N-k$, and the second sheet answers are trustworthy for small $(N-k)/N$.

To test whether the results in equation (\ref{sigma-two-poss}) are in agreement with
a discrete eigenvalue density, we have evaluated the expression in equation (\ref{asym1})
for $\rho(\phi)=N^{-1}\sum_a \delta(\phi-\phi_a)$ with $\phi_a$ taken to match the
large-$N$ distribution (\ref{dist1}) at some finite $N$.  Fig. \ref{numerics1}
shows the results.  For larger values of $\lambda$ ($\lambda\geq e=2.718\ldots$),
the first sheet result dominates for $k/N<1/2$ and the second sheet result dominates
for $k/N>1/2$, with a phase transition at $k/N = 1/2$.  The existence of the phase transition is well
supported by the computation with a discrete density, especially at larger
values at $\lambda$ where the phase transition is sharper.
For  $2<\lambda\leq e$, it would
appear that there are two more phase transitions as the saddle points exchange positions again.
It is clear from the discrete results, however, that these are an artifact of the continuum
approximation.  As described above, we cannot trust the green curve for larger values of $k/N$,
or the blue curve for smaller values of $k/N$.

\begin{figure}
\includegraphics[scale=0.4]{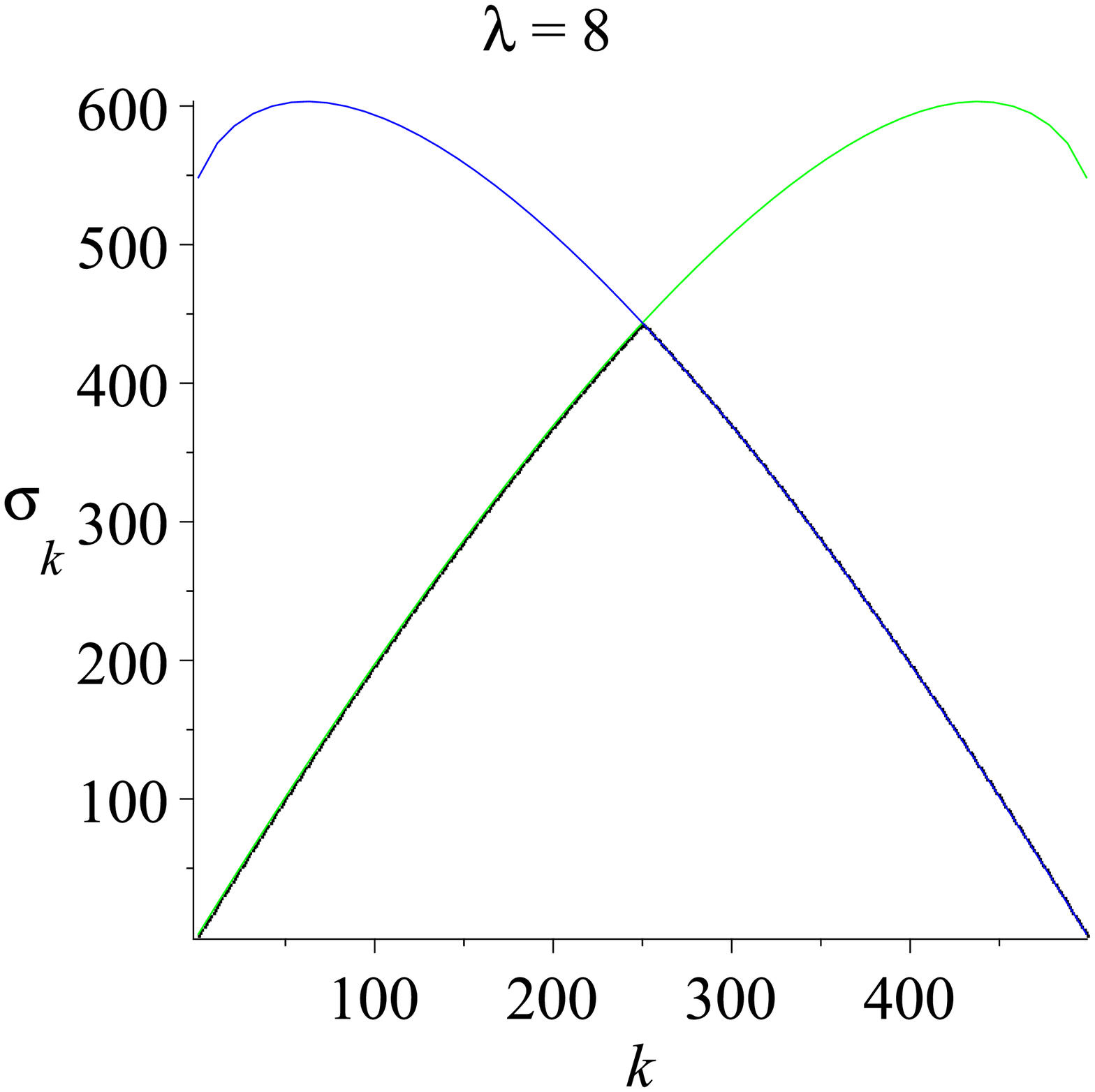}
~~~
\includegraphics[scale=0.4]{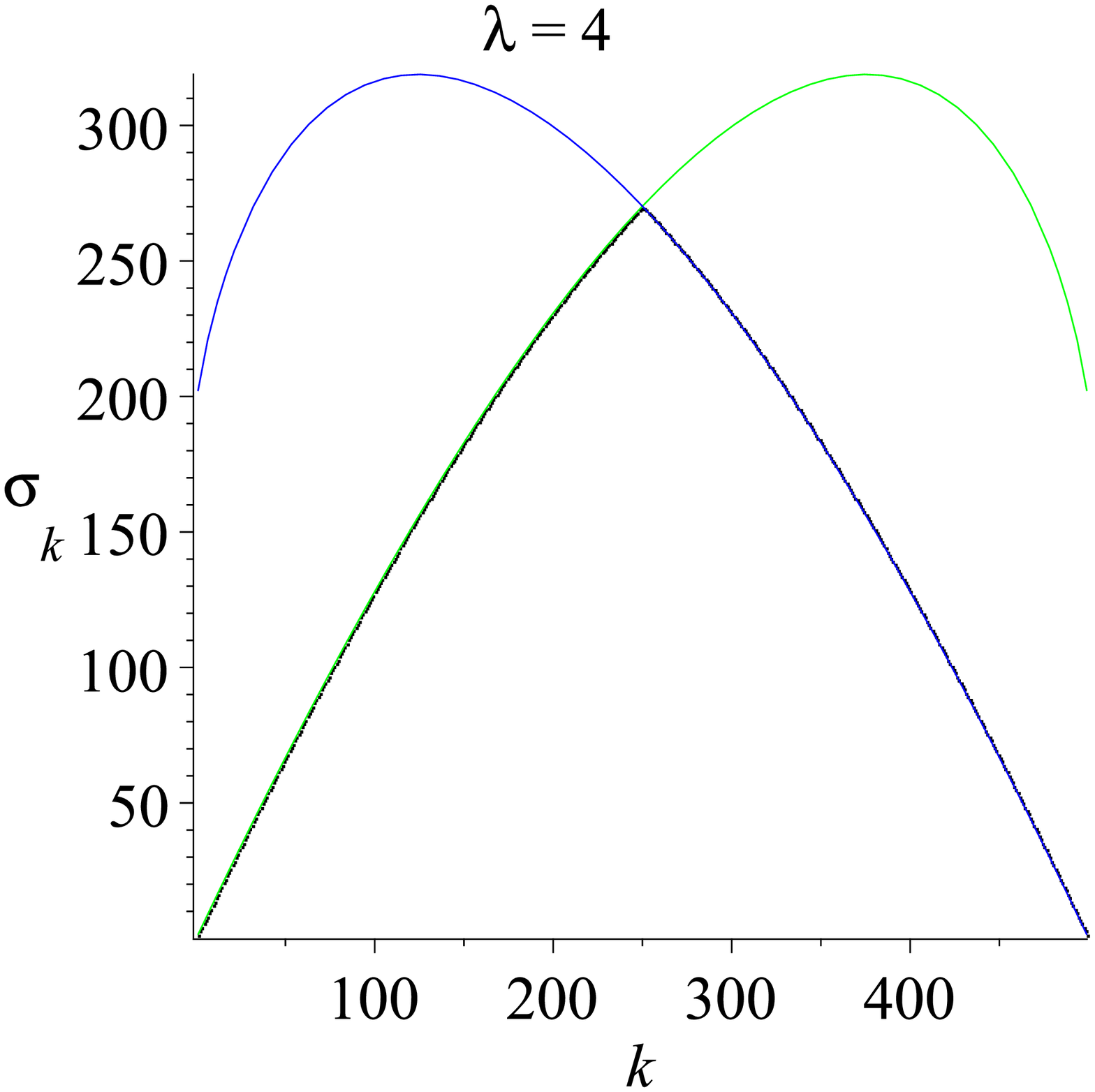}
\\ \\ \\
\includegraphics[scale=0.4]{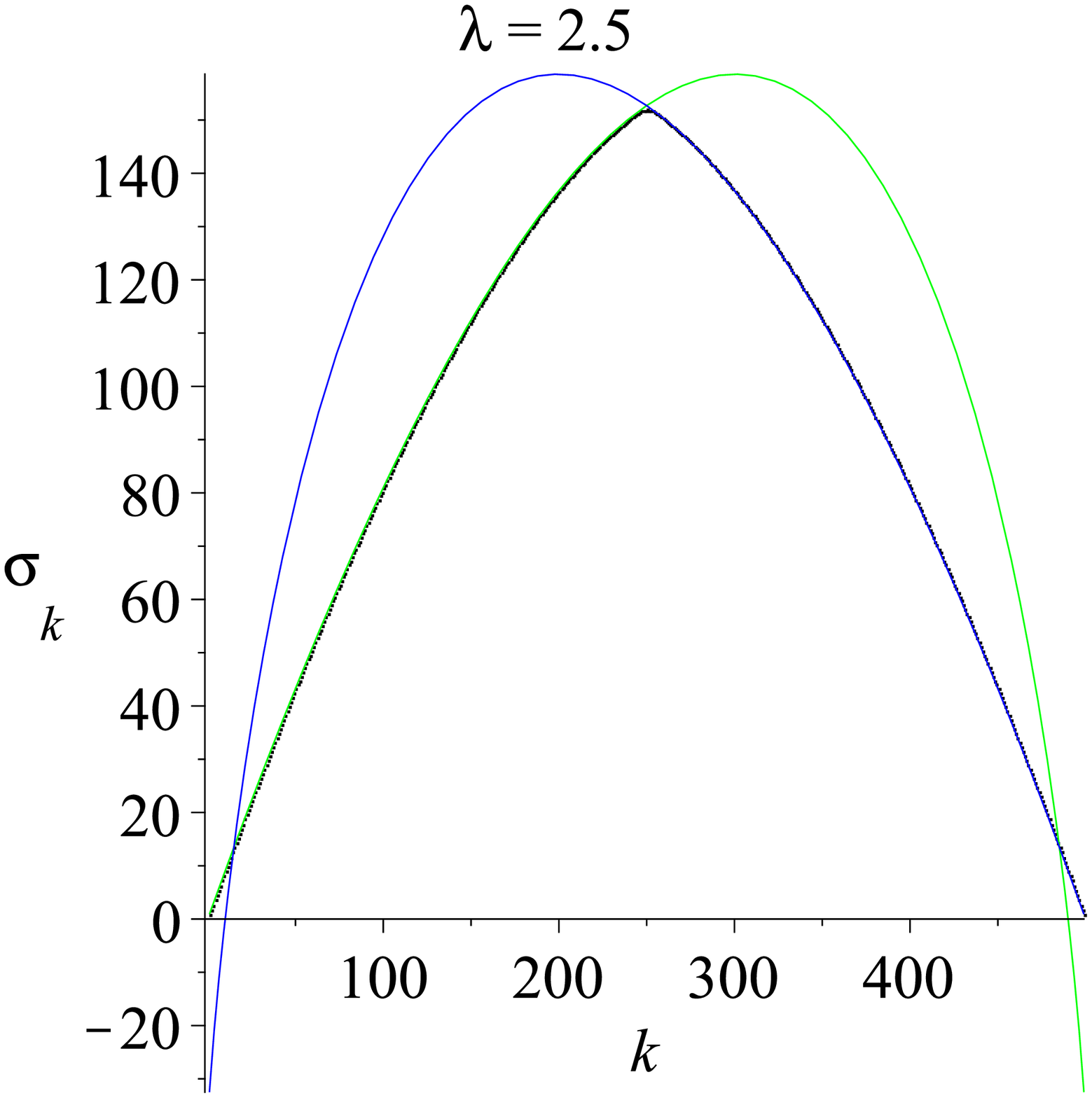}
~~~
\includegraphics[scale=0.4]{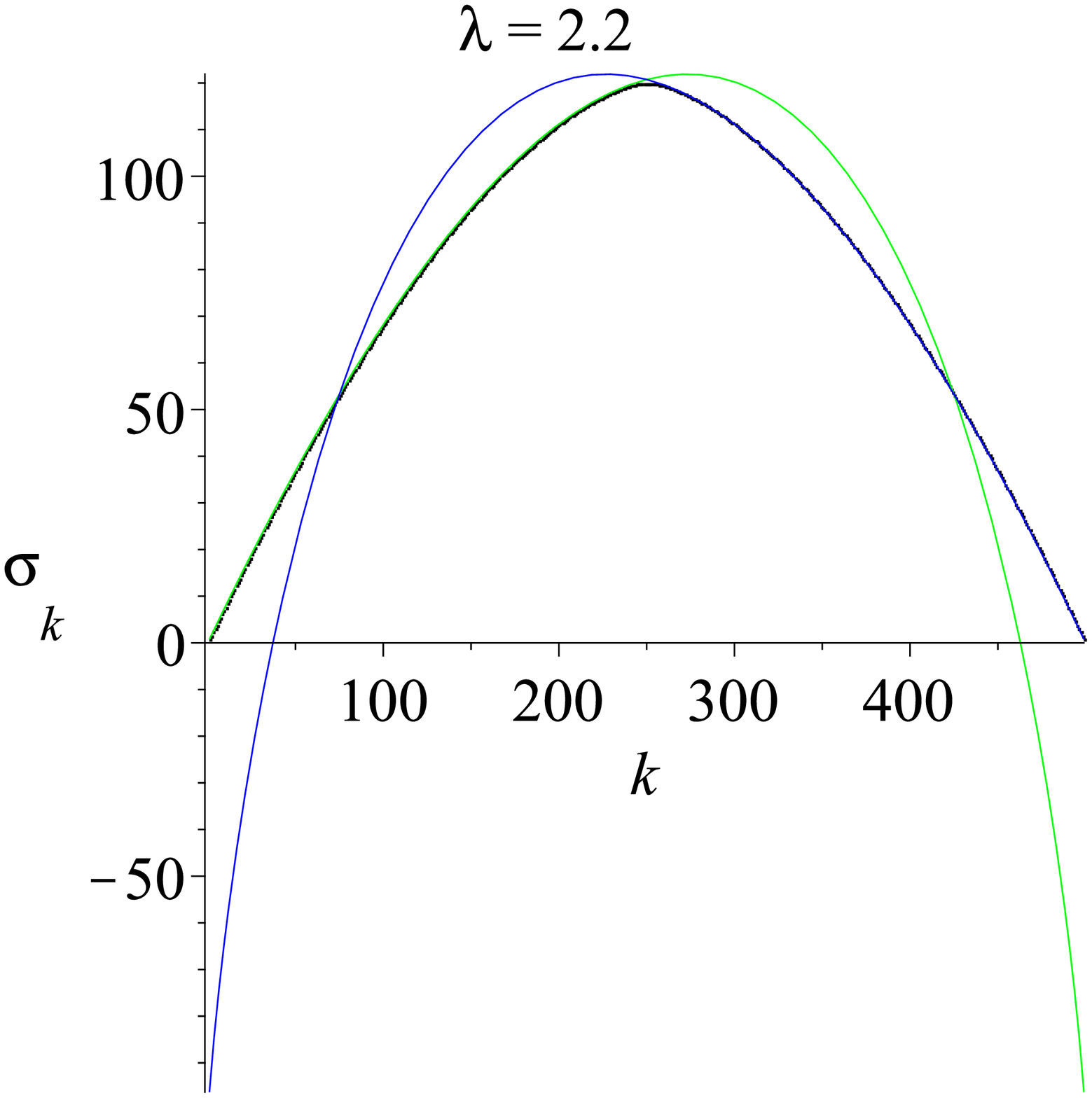}
\caption{$\sigma_k$ as a function of $k$ for $N=500$.
The green and blue solid lines are obtained from the first and second
terms in equation (\ref{sigma-two-poss}) respectively, while
discrete points (black) are computed directly from equation (\ref{asym1}) with
a discrete density $\rho(\phi)$.}
\label{numerics1}
\end{figure}

In summary, $\sigma_k$ depends on whether $\tfrac{k}{N}$ is greater than or less than $\tfrac{1}{2}$,
\be
\label{asym11}
\sigma_k = \left \{ \begin{array}{ll}
   -k-(N-k)\ln(1-\tfrac{k}{N})+k\ln\lambda & \tfrac{k}{N}\leq\tfrac{1}{2}~,~\lambda>2
\\ 
  -(N-k)+k\ln\tfrac{k}{N}+(N-k)\ln\lambda   &\tfrac{k}{N}\geq\tfrac{1}{2}~,~\lambda>2
\end{array} \right .
\ee

These are accurate to the leading order in large $N$, ${\cal O}(N)$, when $k\sim N$ and we
expect that there are corrections of order one, ${\cal O}(N^0)$.
The result  has an explicit $k\to N-k$ duality which is obtained
by exchanging the saddle points. This exchange of saddle points is a first
order phase transition (the first derivative by $\tfrac{k}{N}$ is discontinuous at $\tfrac{k}{N}
=\tfrac{1}{2}$).

We recall that for a single box fundamental Wilson loop in the strong coupling phase of the Gross-Witten model \cite{Gross:1980he}, $\sigma_1 = \ln\lambda$. In the top expression in (\ref{asym11}),
we can clearly identify the contribution from the energies of the $k$ fundamental strings,
$k\sigma_1  = k\ln\lambda$.  We can also confirm from
the remaining contributions in (\ref{asym11}) that the $k$-string tension is indeed always less, that is $\sigma_k \leq k\sigma_1$, as expected for a stable $k$-string.  In fact, for small $\tfrac{k}{N}$,
\begin{equation}
\sigma_k= k\sigma_1-\frac{k^2}{2N}
 +\ldots~.
\end{equation}
As in the continuum string tension, the correction is of order $1/N$, rather than $1/N^2$.

It is interesting to compare the string tensions in (\ref{asym11})  with those in the symmetric
representation \cite{Grignani:2009ua}
\begin{align}
\sigma_{{\cal S}_k} = (N+k)\ln(1+\tfrac{k}{N}) + k\ln\lambda  -k  ~,~\lambda<2~.
 \label{sym11}
\end{align}
The symmetric string tension is always larger than the tension of $k$ single strings,
$\sigma_{{\cal S}_k}\geq k\sigma_1\geq\sigma_k$, even for small $\tfrac{k}{N}$ where
$\sigma_{{\cal S}_k}=k\sigma_1 +\frac{k^2}{2N} +\ldots$.
This is consistent with the idea that representations
other than the antisymmetric one should have higher energies. Also, as expected, the symmetric representation
tension is not identical to the antisymmetric one, as there are no dynamical gluons to implement screening. In this
representation, the strings seem to repel each other.

Now let us consider the weakly coupled, gapped phase with eigenvalue distribution (\ref{dist2}).  The string tension is
obtained by integrating (\ref{sym4}) to get
the exponent
\bear
&& -\int d\phi\rho(\phi)\ln\left(1+\hat te^{i\phi}\right)~+~\tfrac{k}{N}\ln\hat t = \\ \nonumber
&&-\frac{(\hat t+1)\left( \hat t+1 - \sqrt{(\hat t+1)^2-2\lambda\hat t}\right)}  {2\lambda \hat t}
+ \ln  \frac {\hat t+1 - \sqrt{(\hat t+1)^2-2\lambda\hat t} } {\lambda\hat t}
  +\frac{1}{2}+\frac{k}{N}\ln\hat t~,
\eear
where $\hat t$ is adjusted to be an extremum of this expression.
Setting the derivative of the above expression by $\hat t$
to zero yields an equation that is identical to the result of taking
the integral over the eigenvalue density (\ref{dist2}) in the saddle-point equation (\ref{sym4}),
\begin{align}
\tfrac{k}{N}-\tfrac{1}{2}= (\hat t-1)\frac{\hat t+1 -
\sqrt{  (\hat t+1)^2-2\lambda \hat t }}{2\lambda\hat t} ~~,~~\lambda\leq 2~.
\end{align}
The position of the cut singularity in the square root is determined by the cut singularity
of the resolvent in the saddle-point equation (\ref{sym4}). The cut is  on the unit circle, antipodal to the region
 where the eigenvalue density has support.  The cut crosses the negative real axis at $\hat t=-1$. The sign of the square root
has been adjusted so that it has the right behavior at $t\sim 0$. As we follow the negative real axis, the sign must
 flip at $t=-1$ so that the expression  has the correct large (negative) $t$ behavior.  There are two solutions
\begin{align}
\hat t_\pm = \frac{ 1  - \tfrac{\lambda}{4}\left(1-2\tfrac{k}{N}\right)^2}
{2  \left(1-\tfrac{k}{N}\right)}
\pm
 \frac{  1-2\tfrac{k}{N} }
{2 \left(1-\tfrac{k}{N}\right)}\sqrt{\tfrac{\lambda^2}{16}\left(1-2\tfrac{k}{N}\right)^2+ 1-\tfrac{\lambda}{2}}~.
\label{gappedt}\end{align}
The argument of the square root in this expression is positive in the domain of parameters of interest.
Both roots are non-negative real numbers which can vanish only when $\tfrac{k}{N}=0$ or $\tfrac{k}{N}=1$.
They are related by
\begin{equation}
t_+t_-=\frac{\tfrac{k}{N}}{1-\tfrac{k}{N}}
~~,~~
t_-\left(\tfrac{k}{N}\right)=\frac{1}{t_+\left(1-\tfrac{k}{N}\right)}~.
\label{gappedtm}\end{equation}
Note that this is similar to the relationship in (\ref{sheet11}) and (\ref{sheet22}).
For real values of the parameters $\lambda$ and $\tfrac{k}{N}$, the argument of the square root
in (\ref{gappedt}) is always positive, so this square root is unambiguously defined in the
entire region of interest.  It does have two branches which can both be confirmed to
satisfy the saddle-point
equation.  It's the smaller root,  $t_-$,  which lives on the first sheet and
goes to zero as $\tfrac{k}{N} \rightarrow 0$, that
will turn out to be the correct one to use based on the discrete density calculation.

We can substitute $t_-$ into (\ref{gappedt}), taking into account the normalization by adding $\ln{\rm dim}{\cal A}_k
=\ln(N!/k!(N-k)!)$  to the result, to get the tension
\bear
\frac{1}{N}\sigma_k &=&-\tfrac{k}{N}\ln\tfrac{k}{N}-(1-\tfrac{k}{N})\ln(1-\tfrac{k}{N})-\frac{2}{\lambda}\left[ 1-\tfrac{\lambda}{4} - \sqrt{ 1-\tfrac{\lambda}{2}
+\tfrac{\lambda^2}{16}(1-2\tfrac{k}{N})^2}\right] \nonumber \\
&+& \ln\frac{2}{\lambda}\left[ 1+\tfrac{\lambda}{4}(1-2\tfrac{k}{N})-\sqrt{ 1-\tfrac{\lambda}{2}
+\tfrac{\lambda^2}{16}(1-2\tfrac{k}{N})^2}\right] \nonumber \\
&+&\frac{k}{N}\ln\frac{ 1-\tfrac{\lambda}{4}(1-2\tfrac{k}{N})^2 - (1-2\tfrac{k}{N})\sqrt{ 1-\tfrac{\lambda}{2}
+\tfrac{\lambda^2}{16}(1-2\tfrac{k}{N})^2}}{2(1-\tfrac{k}{N})}~.
\label{gappedtension}\eear
Similar expression can be obtained starting with the other root, $t_+$.
Fig. (\ref{numerics2}) summarizes the results: even though the second sheet
saddle point is smaller, it is the first sheet result stated above which
agrees with a discrete eigenvalue density computation.

\begin{figure}
\includegraphics[scale=0.4]{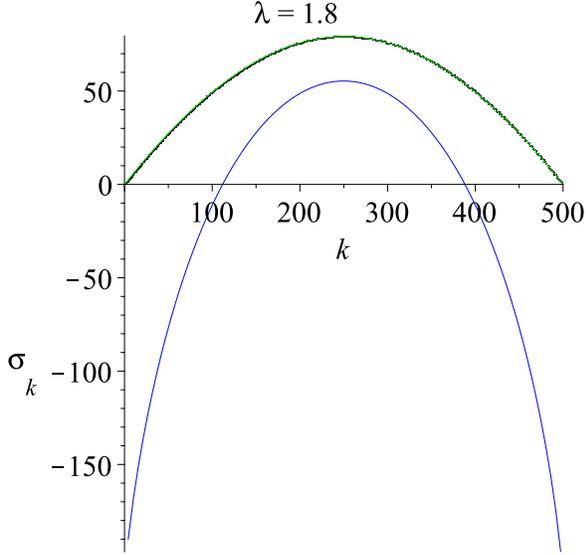}
\caption{$\sigma_k$ as a function of $k$ for $N=500$ in the
weakly coupled phase.
The green and blue solid lines correspond to equation \ref{gappedtension}
and a similar expression obtained for the other sheet of the
saddle point equation.  The green line represents our result
for the $k$-string tension in the semi-circle distribution.
Discrete points (black) are computed directly from equation (\ref{asym1})
using a discrete eigenvalue density.}
\label{numerics2}
\end{figure}

Though it is not immediately obvious from expression
(\ref{gappedtension}), the tension is symmetric under replacement of
$\tfrac{k}{N}$ by $1-\tfrac{k}{N}$.  This is clearly seen in
Fig. \ref{numerics2}.

For small $\tfrac{k}{N}$,
\begin{equation}
\sigma_k = k~\ln\frac{1}{1-\tfrac{\lambda}{4}}~-~
\frac{k^2}{ N}~\frac{\lambda(8-3\lambda)} { 2 (4-\lambda)^2}
~+~{\cal O}\left(\frac{k^3}{N^2}\right)~,
\label{gappedexpansion}
\end{equation}
where, as expected, the string tension for a $k=1$ string in the weak coupling phase is $\sigma_1=\ln\frac{1}{1-\tfrac{\lambda}{4}}$ \cite{Gross:1980he}. As in the strong coupling phase, the interaction
term is attractive and is suppressed by one factor of $1/N$. We can also confirm that the interaction is attractive for all
values of $0<\tfrac{k}{N}<1$ and for all $0\leq\lambda\leq 2$. 


On the other hand, in the weak coupling phase, the symmetric representation
string tension can be deduced
from the results in Ref.~\cite{Grignani:2009ua}
\bear
\frac{1}{N}\sigma_{{\cal S}_k} &=&-\tfrac{k}{N}\ln\tfrac{k}{N}+(1+\tfrac{k}{N})\ln(1+\tfrac{k}{N})+\frac{2}{\lambda}\left[ 1-\tfrac{\lambda}{4} - \sqrt{ 1-\tfrac{\lambda}{2}
+\tfrac{\lambda^2}{16}(1+2\tfrac{k}{N})^2}\right] \nonumber \\
&-& \ln\frac{2}{\lambda}\left[ 1+\tfrac{\lambda}{4}(1+2\tfrac{k}{N})-\sqrt{ 1-\tfrac{\lambda}{2}
+\tfrac{\lambda^2}{16}(1+2\tfrac{k}{N})^2}\right] \nonumber \\
 &+&\frac{k}{N}\ln\frac{ -1+\tfrac{\lambda}{4}(1+2\tfrac{k}{N})^2
+ (1+2\tfrac{k}{N})\sqrt{ 1-\tfrac{\lambda}{2}
+\tfrac{\lambda^2}{16}(1+2\tfrac{k}{N})^2}}{2(1+\tfrac{k}{N})}~.
\label{gappedsymmetrictension}\eear
By considering the small $\tfrac{k}{N}$ limit,
$\sigma_{{\cal S}_k} = k\ln\frac{1}{1-\tfrac{\lambda}{4}}
~+~
\frac{k^2}{ N}~\frac{\lambda(8-3\lambda)}{2 (4-\lambda)^2}
~+~{\cal O}\left(\frac{k^3}{N^2}\right)$,
we see that,  in the symmetric representation, the interaction is repulsive. It is possible to show that it is
is repulsive for all values of $\tfrac{k}{N}>0$ and for all values of $\lambda\leq 2$. 

In conclusion, we have found explicit formulae for the $k$-string tension in both the weak and strong coupling
phases of 2 dimensional lattice Yang-Mills theory. These formulae differ from the continuum result, which is the
quadratic Casimir of the representation.  They nevertheless share the property that the interaction between elementary
strings is attractive when the quark sources are in the antisymmetric representation and repulsive when they are in
the symmetric representation. Furthermore, in the weak coupling phase the $k\to N-k$ duality is realized by a first order
phase transition at $k=\tfrac{N}{2}$. An interesting check of the validity of these results is to examine the continuum limit.
The weak coupling phase has a good continuum limit as $\lambda\to 0$.  In that limit, we should send the lattice spacing to zero, $a\to 0$ while holding the string tension fixed.  If we require that the $k=1$ string tension in inverse distance units
\begin{equation}
\sigma_1=\frac{1}{a^2}\ln\frac{4}{4-\lambda}
\end{equation}
remains finite, we must tune the coupling constant as $\lambda=4-4e^{-\sigma_1a^2}$. Using this in the
$k$-string tension  (\ref{gappedtension}), we obtain
\begin{equation}
\sigma_k  = k\left(1-\tfrac{k}{N}\right)\sigma_1
~+~ \frac {k^2}{3N} \left ( \left (1 - \tfrac{k}{N}\right) a^2 \sigma_1 \right )^2
\sigma_1
~+~ {\cal O}(a^6)~.
\end{equation}
Similarly, for the symmetric representation in (\ref{gappedsymmetrictension}), the continuum limit yields
\begin{equation}
\sigma_{{\cal S}_k} = k\left(1+\tfrac{k}{N}\right)\sigma_1
~-~ \frac {k^2}{3N} \left ( \left (1 + \tfrac{k}{N}\right) a^2 \sigma_1 \right )^2
\sigma_1
~+~ {\cal O}(a^6)~.
\end{equation}
The leading terms are in agreement with the Casimir scaling (\ref{continuumformula})
which is known to be the exact behavior of continuum 2-dimensional Yang-Mills theory.

\vskip .5cm
\noindent
The authors acknowledge hospitality of the Galileo Galilei Institute,
Khavli Institute for Theoretical Physics, Santa Barbara, the
Aspen Center for Physics and the Perimeter Institute.  This work is
supported in part by NSERC of Canada.

\end{document}